\documentclass[11pt]{article}
\usepackage[utf8]{inputenc}
\usepackage{geometry}
\usepackage{setspace}
\usepackage{hyperref}
\usepackage{footmisc}
\usepackage{url}
\title{The Philosophic Turn for AI Agents:\\ Replacing centralized digital rhetoric with decentralized truth-seeking\\ \textit{Penultimate Draft}}
\author{Philipp Koralus\\ HAI Lab\\ Institute for Ethics in AI\\ University of Oxford}
\date{24 April 2025}

\begin{document}
\maketitle

\begin{abstract}
In the face of rapidly advancing AI technology, individuals will increasingly rely on AI agents to navigate life's growing complexities, raising critical concerns about maintaining both human agency and autonomy. This paper addresses a fundamental dilemma posed by AI decision-support systems: the risk of either becoming overwhelmed by complex decisions, thus losing agency, or having autonomy compromised by externally controlled choice architectures reminiscent of ``nudging'' practices. While the ``nudge'' framework, based on the use of choice-framing to guide individuals toward presumed beneficial outcomes, initially appeared to preserve liberty, at AI-driven scale, it threatens to erode autonomy. To counteract this risk, the paper proposes a philosophic turn in AI design. AI should be constructed to facilitate decentralized truth-seeking and open-ended inquiry, mirroring the Socratic method of philosophical dialogue. By promoting individual and collective adaptive learning, such AI systems would empower users to maintain control over their judgments, augmenting their agency without undermining autonomy. The paper concludes by outlining essential features for autonomy-preserving AI systems, sketching a path toward AI systems that enhance human judgment rather than undermine it.
\end{abstract}

\section{Introduction}

Modernity has greatly expanded opportunities for individuals. It has also vastly increased the burden of questions that individuals need to take on board for opportunities to be realized. With increasing capabilities of AI and their impact on how we live and work, this trend stands to dramatically accelerate. It will increasingly be the case that effectively operating in the modern world will require the assistance of AI agents that help us make choices and help us identify what choices and questions we should consider.

Against this background, we face an apparent dilemma that mirrors the ancient image of Scylla and Charybdis. On the one side lies the risk of losing agency: as the complexities of modern life keep compounding and objectively available choices and the stakes involved in negotiating them keep increasing, individuals with their limited cognitive capacities are threatened with becoming ineffective. If we are confronted with the availability of systems that can make us dramatically better at choosing the means to our ends, not using them may ultimately make us lose faith in our agency altogether, particularly if we are routinely in competition with others who happily adapt such systems. But this is not the only peril. On the other side lies the risk of losing autonomy. As our choices become routinely conditioned by automated choice framing and recommender systems, the operative priorities that determine where we end up are less and less originating in ourselves. Instead of enjoying self-rule, we threaten to become ruled by technology, through a kind of ``autocomplete for life''\footnote{Brendan McCord (2024). ``What can Aristotle teach us about AI?''. American Optimist. Ep 105, 14/12/2024.}. In the long run, agency without autonomy is empty, and autonomy without agency is impotent. Worse still, the absence of either tends to corrode the other.

The hazards are real but we cannot turn back. AI is both too consequential and will become too widely embedded to reject. The question is how to navigate the treacherous passage between loss of agency and loss of autonomy. What does it take to design AI agents as products that bolster our ability to act decisively and remain self-governing in the face of new complexities? How can we avoid becoming passive instruments of digital architecture while still reaping its tremendous benefits? In this paper, I will argue that to answer these questions well, we need to distill the essence of philosophical practice into the design of AI products along lines I will outline in this paper. Moreover, we need to pay close attention to the relationship between truth-seeking and decentralization. This perspective goes against a widespread focus on hyper-centralization that is found both in the discourse about AI technological development itself and in discourse about how its risks and ethical dimensions should be approached.\footnote{For recent synoptic discussions heavy on centralization, see: Kissinger, Henry A., Eric Schmidt, and Craig Mundie. \textit{Genesis: Artificial Intelligence, Hope, and the Human Spirit}. New York: Little, Brown and Company, 2024; Suleyman, Mustafa, and Michael Bhaskar. \textit{The Coming Wave: Technology, Power, and the Twenty-first Century's Greatest Dilemma}. New York: Crown Publishing Group, 2023.}

It is tempting to think that there is in fact no dilemma at all, because we might just task AI with improving our well-being in some reasonably liberally minded way, which would then simply subsume what is worth having in agency and autonomy. To address the pitfalls of this view, it is worth considering a version of it that predates the development of AI: the ``nudge'' approach to influencing choice in service of welfare that had its heyday in the first quarter of the 21st century. This approach raises many of the same issues we face in the context of reliance on AI agents, albeit writ small in a way that makes some of the challenges easier to diagnose. I will then zoom out and consider consequences in the context of the new capabilities afforded by AI.

Proponents lauded the nudge approach as liberal paternalism, using minimal interventions like default enrollments or psychologically informed choice framing to guide individuals toward better outcomes.\footnote{Thaler, R. H., \& Sunstein, C. R. (2008). \textit{Nudge: Improving decisions about health, wealth, and happiness}. New Haven, CT: Yale University Press.} The nudge approach could be seen as an attempt to resolve our dilemma, at an earlier stage of technological development: individuals are seen to make sub-optimal choices, and scientifically informed interventions are made as a partial remedy, in a way that (so the hope goes) does not impinge on our freedom.  The philosophy of nudges also foreshadows the temptations of being ruled by a new kind of technology; temptations that are rapidly growing with AI. If it could be made to work very broadly, a naïve philosophy of nudging would be the apotheosis of technocratic government: guiding the masses through techne alone with no need for physical force or even traditional propaganda. Augmented with AI, this approach promises the utopia of autocomplete for life.

I will argue that when fused with large-scale AI systems, nudge philosophy risks turning into a subtle but sweeping form of digital rhetoric, one that can invisibly shape the collective mind, inhibit the decentralized adaptive learning that underlies human truth-seeking, and ultimately undermine human autonomy. At AI-driven scale, nudging threatens to turn ``soft-paternalism'' into a kind of soft totalitarianism, to the detriment of long-term vitality of human reason. The key problem is to instead chart a course for AI agents that can help us maintain effective agency in an ever more complex world, while allowing us to maintain autonomy.

This paper will conclude by presenting a paradigm for how we can chart such a course, arguing for a philosophic turn in AI agents. This marks a shift from the rhetorical school of designing ``choice architectures'' to cultivating systems that foster decentralized truth-seeking. Rather than nudging users into predetermined pathways, we could build AI to ask catalytic questions while keeping complexity at bay, facilitate open-ended inquiry, and help preserve autonomy while augmenting agency. Such a system could help us adapt to changing circumstances, not by minimizing the user's role, but by elevating it. Personal AI systems could help us navigate our complex future, while empowering us to remain the authors of our own judgments. I will outline what I take to be some key necessary features an AI product would need to have to deliver these goods. I conclude that the right design principles for AI systems could guide us through our dilemma, preserving both autonomy and agency.

\section{Background on nudges}

Nudges, to a first approximation, are interventions that change people's behavior without changing the range of choices that are objectively available to them. For example, this could involve having an employer enroll their employees in an optional retirement plan by default while allowing them to opt out. This would tend to make it more likely that employees end up on this plan, without forcing them to do so.\footnote{Madrian, B.C., \& Shea, D.F. (2001). ``The Power of Suggestion: Inertia in 401(k) Participation and Savings Behavior.'' \textit{The Quarterly Journal of Economics}, 116(4), 1149--1187.} Another example would be a government policy to make everyone an organ donor by default unless they explicitly opt out.\footnote{Johnson, Eric J., and Daniel Goldstein. ``Do Defaults Save Lives?'' \textit{Science} 302, no. 5649 (2003): 1338--1339.} A different type of nudge involves fitting goods like sugary soda or tobacco with warning labels emphasizing their negative health impact\footnote{Hammond, David, Geoffrey T. Fong, Ron Borland, K. Michael Cummings, Ann McNeill, and Pete Driezen. ``Text and Graphic Warnings on Cigarette Packages: Findings from the International Tobacco Control Four Country Study.'' \textit{American Journal of Preventive Medicine} 32, no. 3 (2007): 202--209.}, or distributing communications designed to increase adherence to laws by talking about how others already follow the rules.\footnote{Hallsworth, Michael, John A. List, Robert D. Metcalfe, and Ivo Vlaev. ``The Behavioralist as Tax Collector: Using Natural Field Experiments to Enhance Tax Compliance.'' \textit{Journal of Public Economics} 148 (2017): 14--31.} Many governments and institutions have adopted nudges of one form or another. The tantalizing promise of nudges as a tool of governance lies in the possibility of promoting human welfare without using the kind of coercion that abridges individual freedom of action, and without incurring significant cost.

The concept of nudges becomes operational through influencing or designing the ``choice architecture'' that provides the opportunity for people to exercise their choices. In the case of default enrollment in a retirement plan, this approach would normally involve designing a questionnaire, that asks if people wish to opt out of the plan rather than asking them whether to opt in. Since, objectively, no option has been taken off the table, nobody has been coerced. Moreover, assuming that being enrolled in such a plan is on average making lives better, we have increased welfare. The most ambitious vision for nudges, and for the so-called ``behavioral insights'' teams administering them, would be to find as many opportunities as possible to have this kind of impact. This would then give us a vision of liberal paternalism; ``liberal'' in the sense that it does not abrogate freedom (conceived as non-interference with our objectively available choices), and ``paternalistic'' in the sense that it has an impact in a way that is guided by a substantive view of what the right choices are for people.

Returning to the default retirement plan example, we might observe that there will be people who end up sticking with a default simply because they never bother to even consider the choice implicit in it. In the retirement plan case, we could end up staying in the default plan simply because we never look at the paperwork that asks if that is how we would like to proceed. We could then complain that these people now only get their full choice reinstated if they ``pay up'' in the form of the opportunity cost of considering the form. This would then lessen the claim to nudges being entirely liberal. However, nudges can be made more sophisticated. This is where modern cognitive psychology and behavioral economics come in.

It is pervasively documented in the behavioral science literature that whether a binary choice is presented in a frame that asks whether an option is rejected tends to yield different average outcomes compared to a frame that asks whether an option is accepted, even if the context ensures that the objective choice situation is the same.\footnote{Shafir, Eldar. ``Choosing Versus Rejecting: Why Some Options Are Both Better and Worse Than Others.'' \textit{Memory \& Cognition} 21, no. 4 (1993): 546--556; Park, C. Whan, Sung Youl Jun, and Deborah J. MacInnis. ``Choosing What I Want Versus Rejecting What I Do Not Want: An Application of Decision Framing to Product Option Choice Decisions.'' \textit{Journal of Marketing Research} 37, no. 2 (2000): 187--202.} Thus, even if we insist on every individual responding to the questionnaire, thereby removing any asymmetry in transaction costs for accepting versus rejecting the retirement plan, the frame will still tend to affect the frequency of objective choice outcomes. Framing effects and the fact that they are predictable in the aggregate create the possibility of what you might call ``pure nudges'' that plausibly fit the narrow notion of liberal paternalism described above, and that are guided by a combination of behavioral science and substantive views of welfare.

We might then be tempted to walk away with what you might call the ``naïve view'' of nudges: Nudge theory provides a general tool that can be deployed in a purely technical manner to further welfare without abrogating liberty. In his 2016 book, \textit{The Ethics of Influence}, Sunstein declares that, ``we live in an age of psychology and behavioral economics''\footnote{Sunstein, Cass R. \textit{The Ethics of Influence: Government in the Age of Behavioral Science}. New York: Cambridge University Press, 2016, p. 1}, with governments relying on psychologists and behavioral insights teams to inform liberal policy for the greater good. Whether or not that statement was accurate close to a decade ago, we now have firmly moved into the age of AI. As I will argue further below, this amplifies serious worries one might have about the framework of nudges.

\section{Are nudges inevitable?}

Sunstein sees the framework of nudges as effectively unassailable for reasons that are summed up in the following paragraph:

\begin{quote}
I shall argue that at least if they are taken in general or in the abstract, the ethical objections to nudging lack force, and for two different reasons. First, both choice architecture and nudges are inevitable, and it is therefore pointless to wish them away. Second, many nudges, and many forms of choice architecture, are defensible and even mandatory on ethical grounds, whether we care about welfare, autonomy, dignity, self-government, or some other value. But it remains true that some nudges, and some forms of choice architecture, are unacceptable. The most obvious and important reason is that they have illicit ends. They might be intended, for example, to entrench the current government, or to help powerful private groups, or to advantage certain racial or religious majorities. But even when the ends are legitimate, public officials owe citizens a duty of transparency, and they should avoid manipulation.\footnote{Ibid., p.12}
\end{quote}

On this view, objections to nudges on general grounds are flawed for some of the following reasons: (1) nudges inevitably arise since many or all choices are made relative to a choice architecture, (2) concerns about various values like welfare make many nudges ethically permissible and even obligatory, (3) the nudge theorist is not committed to every nudge being acceptable, notably ruling out those that serve ``illicit ends'', or that help certain groups or individuals, and further ruling out manipulation, while demanding transparency.

(3) seems entirely reasonable but it does mean that we cannot maintain the attractive naïve view of nudges outlined in the previous section. We cannot permissibly deploy techniques based on framing effects on a purely technical basis determined by behavioral science, even if we are doing so in service of welfare. The permissibility of each case must be decided in a case-by-case manner that involves much broader context and political questions. Instead of saying that all nudges are acceptable, the nudge theorist can at least plausibly say that there exist cases in which nudges are acceptable to deploy in the public interest. However, they do not become permissible simply because they take the form of nudges.

The idea (2) that some nudges may be obligatory is interesting in this context. In cases in which we might deem nudges to be morally obligatory, would we also deem more coercive policies to be obligatory? For nudges to have a distinctive status, we would have to convince ourselves that the answer is often ``no''.

The ``master argument'' appears to be that all choices are made relative to some choice frame, and hence ``both choice architecture and nudges are inevitable''\footnote{Ibid., p.12}. I will focus discussion on the case of defaults, since Sunstein takes those cases to be the ``most obvious and important nudges.''\footnote{Ibid., p.26} For example, in the case of the pension plan example, we only have three possibilities: (A) enrolment in the plan is the default and people can opt out (B) not being enrolled in the plan is the default and people can opt in (C) employees are presented with a forced choice and may not proceed without deciding to either enroll in the plan or not. Both (A) and (B) can be seen as nudges. (C) does not qualify as a default nudge, assuming the wording is suitably symmetrical between options, but it could be argued that it introduces an ``illiberal'' coercive element since the employee is not able to proceed with whatever else they might have wanted to do without first working through the questionnaire. Moreover, (C) could still accidentally become a nudge if we don't get our survey wording right. For example, even if we present the choice both in terms of ``accepting'' the retirement plan and in terms of ``rejecting'' it, we could still potentially get an order effect that could subtly encourage one choice over another.

Context and framing effects, while systematic, are not so pervasive that every choice can be pushed around through objectively equivalent reframing. A superficial reading of popular overviews of fallacies and framing effects might give the impression that framing effects are everywhere and easy to find once we throw off the fetters of the ideology of classical ideal rationality.\footnote{Kahneman, Daniel. \textit{Thinking, Fast and Slow}. New York: Farrar, Straus and Giroux, 2011; Ariely, Dan. \textit{Predictably Irrational: The Hidden Forces That Shape Our Decisions}. New York: Harper, 2008.} In practice, it is a hard-won achievement (worthy of the occasional Nobel prize) to find systematic framing effects that yield significant differences. Moreover, certain framing effects like the famous endowment effect have been shown to diminish with trading experience.\footnote{List, John A. ``Does Market Experience Eliminate Market Anomalies?'' \textit{Quarterly Journal of Economics} 118, no. 1 (2003): 41--71.} It appears that the difference experience makes even potentially has an identifiable neural correlate.\footnote{Tong, Lester C. P., Karen J. Ye, Kentaro Asai, Seda Ertac, John A. List, Howard C. Nusbaum, and Ali Hortaçsu. ``Trading Experience Modulates Anterior Insula to Reduce the Endowment Effect.'' \textit{Proceedings of the National Academy of Sciences} 113, no. 33 (2016): 9238--9243.} Some have even argued that no empirical evidence for nudges remains after correcting for publication bias.\footnote{M. Maier, F. Bartoš, T.D. Stanley, D.R. Shanks, A.J.L. Harris, \& E. Wagenmakers, No evidence for nudging after adjusting for publication bias, \textit{Proc. Natl. Acad. Sci. U.S.A.} 119 (31) e2200300119, \url{https://doi.org/10.1073/pnas.2200300119} (2022).} Thus, for many choices, there will be a large class of objectively equivalent frames that yield roughly the same results on average. In those cases, it would not be right to say that framing effects are inevitable.

The nudge theorist might reply that it would still be the case that some framing effects are inevitable in some cases, since some choices have to remain in a frame of defaults. If every choice that people may encounter is given a treatment along the lines of (3) above, this would drastically overtax people's attentional resources. This reply seems entirely plausible and returns us to the problem of modernity from our introduction.

The nudge theorist is not merely arguing that some framing effects are inevitable but that nudges and choice architecture are inevitable. This requires further assumptions. Sunstein thinks that the distinction between actions and omissions is wrong to the point of being ``possibly incoherent''\footnote{Ibid., p. 16}. If we have no distinction between action and omission, then all choice frames that might be connected to framing effects, presumably including all frames involving a default of some sort, are nudges in much the same way as a choice frame purpose-built by a behavioural insights team. However, this way of arguing for the inevitability of nudges is problematic if we are interested in nudges in the context of policy-making. Even if you do not think there is a difference between action and omission at the level of individual moral action, a government still must make that distinction unless it is totalitarian. A liberal government must omit to stop many things that it may not do.

Further assumptions are similarly needed to go from the inevitability of choice frames to the idea that choice architecture is inevitable. Compare say a claim that urban planning is inevitable because people live in cities. While the shape of many or even most cities is significantly the result of a concerted effort at planning, the layout of many old cities is the result of the gradual evolution of the built-up environment in response to various constraints. We could of course use the term urban planning to include these emerging city layouts as a kind of edge case. Then perhaps we could also use the term choice architecture in this wider sense. It is not clear to me what is gained by this choice of language. Perhaps we could also discuss uncharted parts of the Amazon rainforest as instances of garden architecture and argue that garden architecture is inevitable. Sunstein seems to be happy with this kind of usage, since he holds that, ``our choices are often an artifact of an architecture for which no human being may have responsibility -- a sunny day, an unexpected chill, a gust of wind, a steep hill, a full (and romantic) moon.''\footnote{Ibid., p.36}

The point of the foregoing is not to argue that it can never be useful or appropriate to use framing effects or defaults in a way that attempts to encourage particular choices, or that there is no legitimate sense in which using such an approach could be more liberal than coercive methods. However, without further assumptions or unattractive terminological stipulations, we do not get clear shortcuts that make nudge policies more ready for deployment than other types of policies, nor do we get that nudges or choice architecture are inevitable in a substantive sense.

\section{Decentralized adaptive learning}

The considerations in the previous section bring us to what might be called the Hayekian objection to nudges. Sunstein denies that there is a relevant difference between a choice ecosystem that is the product of gradual social evolution, and a choice architecture that has been designed by a planner. Spontaneous orders and intentionally designed choice architecture are on a par, in this regard.\footnote{Ibid., p.38} Moreover, he is sceptical that spontaneous orders have a particularly special status with regard to making good choice architecture (in Sunstein's very broad usage of the term). He remarks that,

``Perhaps traditions and customs are reliable; if they have managed to survive over time, it might be because they are sensible and helpful and really do make people's lives better. Perhaps public officials, or law, can build on traditions and avoid any kind of top-down dictation. Even if traditions are a form of choice architecture, and even if they nudge, they might be trustworthy by virtue of their longevity [\ldots] In my view, the strongest position in favor of spontaneous orders and invisible hands cannot, in the end, be defended, because such orders and such hands do not promote welfare, autonomy, dignity, or self- government. But the minimal point is that a degree of official nudging cannot be avoided. If we are committed to spontaneous orders and invisible hands, we will be committed to a particular role for government, one that will include a specified choice architecture and specified nudges.''\footnote{Ibid., p. 38-41}

We will set aside the question of what the right scope for the role of the state is in general. On all but the most extreme views, the state will have to enforce various measures, beginning with a monopoly on violence but far from ending with it. The more interesting question is whether the choice ecosystem as first encountered by the nudge theorist is a kind of fertile no-man's land where new edifices of choice architecture may be erected with relatively few constraints (e.g. Sunstein's caveats mentioned above about avoiding manipulation, illegitimate aims, and the like). This is a key question that the choice of language favoured by nudge theorists makes difficult to even ask, since, as we saw, for Sunstein even a ``(romantic) moon'' is a kind of choice architecture.

It is worth briefly bringing into view some of the benefits of spontaneous order. Two of the most interesting forms of spontaneous yet complex order that have had the greatest impact on humanity as a whole are science, or the more broadly conceived Wissenschaft, and the market economy. Setting aside extreme views, it is misleading to say that ``spontaneous orders and invisible hands'' do not ``promote welfare, autonomy, dignity, or self- government.'' Nothing has lifted more people out of poverty over the past century than the market economy, which is a foundation for the possibility of widespread welfare, autonomy, dignity, and self-government (even if it does not guarantee those things).\footnote{Deaton, Angus. \textit{The Great Escape: Health, Wealth, and the Origins of Inequality}. Princeton: Princeton University Press, 2013.} These improvements in turn were powered by technological progress driven by science. We do not have to take the position that processes driven by spontaneous order are sufficient for the level and distribution of welfare and other goods that we would like, in order to acknowledge that all practical evidence points to such processes being necessary for it. It is misleading to evaluate the relative importance of spontaneous orders as the consideration that, in Sunstein's words, ``perhaps traditions and customs are reliable'', as if we were discussing the relative merits of wearing trousers or more hardwearing lederhosen to the city.

Science and the market economy are both the result of gradual and decentralized adaptive learning. I will focus on these two examples, though there are many others.\footnote{Ridley, Matt. \textit{The Evolution of Everything: How New Ideas Emerge}. New York: Harper, 2015.} In the modern economy, prices are for the most part not set by a central planner but are the result of the decentral forces of supply and demand. For goods that allow for reasonably dynamic markets and where there are no specific other special constraints, few doubt that this is the most efficient way to determine prices. Indeed, as economic history shows, it is incredibly difficult to intervene in this process through top-down controls in a way that does not end up reducing aggregate welfare.\footnote{Mankiw, N. Gregory. \textit{Principles of Economics}. 9th ed., Cengage Learning, 2020.} This is a case of decentralized adaptive learning because the price is the result of the aggregate of individual decision-makers making choices and responding to their environment.

In the case of science, we similarly have decentralized adaptive learning. There is no central agency that decides what ``the science'' says. Scientific knowledge is scattered globally among many researchers who largely set their own research agendas. Adaptive learning happens as a result of the efforts of individual teams of researchers and feedback comes from peer review (made possible by overlapping rather than identical expertise) and open discourse, combined with competition for funding and posts.\footnote{It should be noted that a degree of decentralization neither guarantees openness to new ideas nor that the incentive structures driving adaptation will directly advance truth-seeking. For a discussion of relevant failures of decentralized systems see: Herman, Edward S., and Noam Chomsky. \textit{Manufacturing Consent: The Political Economy of the Mass Media}. New York: Pantheon Books, 1988.}

Despite the incredible advances that science and the market economy have produced, the most important strength of both is arguably not the value, however measured, of the current state of economic output or scientific knowledge. In the long run, the mechanism of adaptation is more important than the current state it has produced. The world keeps changing, so we have to keep adapting. The importance of the mechanism of adaptation is particularly salient in science. In the case of science, the core aim is the continuous production of new knowledge rather than merely to be a custodian for a certain fixed body of knowledge. But this aim is not just a kind of greed for ever more knowledge to which we could respond with a kind of therapeutic suggestion (``why can't you just be satisfied for once!?''). It is by being embedded in the process of producing new knowledge that old knowledge maintains a status of robustness, since it is continuously reassessed in light of new discoveries and insights. But without robustness or ``safety'', most contemporary philosophers of knowledge would take it, there is no knowledge at all.\footnote{Ichikawa, Jonathan Jenkins and Matthias Steup, ``The Analysis of Knowledge'', \textit{The Stanford Encyclopedia of Philosophy} (Fall 2024 Edition), Edward N. Zalta \& Uri Nodelman (eds.), \url{https://plato.stanford.edu/archives/fall2024/entries/knowledge-analysis/}} We could plausibly claim that the ongoing decentralized truth-seeking of the scientific community is in fact a necessary condition for any scientific knowledge to continue to count as knowledge. In J.S. Mill's terms, ongoing inquiry is what distinguishes ``living truths'' from ``dead dogma''.\footnote{Mill, John Stuart. \textit{On Liberty}. London: Longman, Roberts \& Green, 1869, Ch. 2} Taking this condition seriously, we might observe that if we were reduced at the civilizational level to a collection of inherited textbooks administered by technical instructors, we might begin to doubt that we can legitimately treat the claims in those books as knowledge, rather than appeals to ancient authority that happen to be correct.

The upshot is then this: some adaptive decentralized processes do have claim to a special status. The most distinctive aspect of that special status is that they provide the most robust and general self-correcting facility we have available, and which is at the foundation of both the most central motors of human welfare and knowledge. Viewed through that lens, a key question in comparing our broad choice ecosystem with specific centrally planned nudges is not whether some particular default is ``correct'', but how it relates the ability of individuals and aggregates of individuals to adapt. The seed corn counts for more than the harvest.

Let us take another look at the paradigmatic example of default enrolment in a retirement plan. From the perspective of adaptive decision making, we could observe that from the early to mid-20th century up until the 1990s it was common in both Europe and the United States for employees to have a defined benefit or final salary pension to look forward to, requiring little or no special decision-making on the part of the employee.\footnote{Seburn, Paul W. 1991. ``Evolution of Employer-Provided Defined Benefit Pensions.'' \textit{Monthly Labor Review} 114 (12): 16--23.} In this world, just taking what you are given without thinking about it too hard is an adaptive response. However, after the financial environment changed and various pension reforms were enacted, this strategy became maladaptive to a changing environment that now required more sustained planning. We could argue that since many of those transformations were themselves the result of centrally made policy choices, introducing default enrolment in a retirement plan against the background of these changing circumstances is as justified as putting up a diversion sign after a planned road closure. However, if we zoom in on the role of adaptation and spontaneous order, we also notice that the previous way retirement schemes worked already had paternalistic tendencies, with the effect of removing certain aspects of long-term financial planning from people's active consideration. This in turn likely made individuals less able to adapt to changing circumstances. This observation would then raise the question of not just whether the new default nudge is in the best interest of individuals as the world currently stands, but what the effect of this default is on people's ability to adapt. For example, default pension plans focus on index funds, since default plans are often intended to be conservative investment strategies with minimal management fees. If these defaults cause investments in index funds to dramatically increase in the aggregate, this could cause distortions. Some economists have worried that the widespread shift from active to passive investments could cause risks to financial stability.\footnote{Anadu, Kenechukwu, Mathias Kruttli, Patrick McCabe, and Emilio Osambela. 2020. ``The Shift from Active to Passive Investing: Potential Risks to Financial Stability?," Finance and Economics Discussion Series 2018-060r1. Washington: Board of Governors of the Federal Reserve System, \url{https://doi.org/10.17016/FEDS.2018.060r1}.} A larger proportion of the economy being tied up in index funds could also make the economy less innovative, which could harm aggregate welfare in the long run. Finally, of course, at least on some trajectories that some experts consider possible, very powerful AI systems belonging to privately owned rather than publicly listed companies could absorb the lion's share of the economy in a way that could undermine index funds.\footnote{Anton Korinek, Jai Vipra, Concentrating intelligence: scaling and market structure in artificial intelligence, \textit{Economic Policy}, Volume 40, Issue 121, January 2025, Pages 225--256, \url{https://doi.org/10.1093/epolic/eiae057}} Now the point here is not to make a prediction or and evaluation of any of these possible risks. The point is to illustrate that many factors, including factors potentially furthered by nudges themselves, could cause the economic environment to again change in a way that makes the current default practice maladaptive on the individual level.

Through the lens of decentralized adaptive learning, it becomes just as important to ask how defaults interact with individual choices in an ever-changing world as it is to ask how such defaults affect individuals today. In addition to asking whether a given intervention in the choice ecosystem is making a given individual or group of individuals worse off we also should ask how this intervention is affecting the ability to learn from and adapt to changing circumstances. It is likely that the widespread deployment of AI systems will tend to increase the pace of change rather than decrease it.

\section{AI and the pitfalls of personalization}

Existing nudge strategies could be criticized as inadequately personalized, where having an average positive benefit can mask that it does not help those who are most in need at all, and that it may even harm some individuals. For example, warning labels on sugary drinks have only a small effect on people who have self-control problems (e.g. the population that is most hoped to benefit from the intervention of having labels), while those who do not have self-control problems are more strongly affected and dislike the labels.\footnote{Allcott H, Cohen D, Morrison W, Taubinsky D. 2022. When do ``nudges" increase social welfare? NBER working paper no. 30740. Available via NBER. \url{https://www.nber.org/system/files/working_papers/w30740/w30740.pdf}. Accessed 25 Mar 2024} Similarly, a coarse-grained default to an index fund retirement plan may be harming some people for whom this option isn't right, even if it may currently be a benefit to people on average. The fact that a nudge intervention increases average welfare does not guarantee that it does not also lower the welfare of significant groups of individuals. In light of such results, Sunstein suggests that AI could allow us to address these problems through personalization, subject to continuing vigilance about the dangers of inserting illegitimate aims, manipulation, and the possibility of AI replicating human biases or inserting new ones.\footnote{Sunstein, Cass R. 2024. ``Choice Engines and Paternalistic AI.'' \textit{Humanities and Social Sciences Communications} 11 (1): Article 34. \url{https://doi.org/10.1057/s41599-024-03428-0}.}

The choice of the sugar warning labels as an example is interesting, since it makes salient the question of who is in control of the AI being consulted, and as whose agent the AI system serves. It is one thing for a health agency to confront me with a warning label when I wish to drink a soda or purchase tobacco products. It is another thing for, say, an agent hired (or purchased) by me to confront me with a gruesome warning that I actively dislike. Who would purchase such an agent privately unless it was the only option allowed? The individualized AI counterparts of many types of nudges that have already been deployed only make sense if we imagine the AI as being in significant part centrally controlled. This is where I suggest the framework of nudges unravels in the age of AI.

We cannot easily defend the stance that defaults ipso facto undermine agency and autonomy, since there may simply be too many choices to practically consider. Choice paralysis in the face of too many decisions is a threat to meaningful agency. The trouble with the ``nudge'' frame is that it paradigmatically is concerned with creating choice architecture in service of nudging people's choices in a direction that is in accord with whatever (hopefully legitimate) ends a choice architect has in mind. In addition to the possibility of fine-grained individualization of nudges, we also get a potential unlimited increase in scope of the kinds of decisions covered. If we imagine people routinely making decisions in a way that is mediated through AI, then the potential areas of life in which nudges could be applied could be unlimited. This presents serious problems.

Let us consider nudge personalization as a supposed advantage. It is worth recalling that one of the biggest scandals with regard to the use of technology in service of manipulative ends, the Cambridge Analytica scandal, precisely centered on mechanisms for highly personalizing how frames are constructed for individuals through targeted communications based on a highly personalized model of those individuals.\footnote{Bakir, Vian. ``Psychological Operations in Digital Political Campaigns: Assessing Cambridge Analytica's Psychographic Profiling and Targeting.'' \textit{Frontiers in Communication} 5 (2020): Article 67. \url{https://doi.org/10.3389/fcomm.2020.00067}.} The problem here is that fine-grained personalization of communications and choice frames makes it considerably harder for the community to hold would-be nudgers and influencers to account, since there is an inherent loss of transparency. Personalization is inherently at odds with maintaining the ability of the community to collectively identify an intervention. If I am shown warning labels about sugar but you are not, then we lack a shared experience on the basis of which we could problematize the relevant policy in shared political discourse. At the limit, if everyone is being nudged in entirely different ways, the only shared experience may be a generic warning that our AI ``may contain nudges in the public interest''. Such warnings would be as practically pointless in giving control to the public as the much-derided EU-style cookie banners. Personalization of central interventions is directly in tension with transparency and accountability, so AI powered personalized nudges are problematic on Sunstein's own account of the limits of nudges.

The increase in scope of areas in which we might have AI-mediated decision making that can be exploited for nudges raises serious problems for the possibility of adaptive decentralized learning that is arguably critical for maintaining a healthy choice ecosystem over time. Particularly if we are considering an AI system that is centrally controlled enough to make nation-wide or industry-wide nudges possible, we risk filtering collective learning through a static template that is not sufficiently able to adapt to local and temporal variation. However personalized, a nudge as a would-be justified intervention is always relative to a model of how the world currently works and how various actions relate to likely outcomes. The more we encounter decisions about how to act through the lens of an AI system that already has such a model baked in, the less we can ultimately learn from contact with reality. For example, we would be less likely to get an opportunity to think through the dimensions of merit of those alternatives that the baked-in model would tend to rule out. Central AI systems are not guaranteed to adaptively learn from new situations in the same way that decentralized spontaneous orders do, and the greater the role of such systems, the less we might find such spontaneous orders emerging with benefit. Trying to solve this problem by having central AI systems design their own nudges relative to an automatically evolving model of the world and value system would in turn entail a potentially dramatic abdication of sovereignty, threatening democratic accountability even more.

Very broad usage of AI agents for decision support could also risk becoming self-reinforcing. We partly discover values and reasons through the process of decision making.\footnote{Chang, Ruth. ``Hard Choices.'' Journal of the American Philosophical Association 3, no. 1 (2017): 1?21. https://doi.org/10.1017/apa.2017.7.?} If we have AI decision aids involved in very wide spaces of decision questions, in contrast with the limited and infrequent nudges of the previous era, they could end up shaping our criteria for evaluating choices in a way that narrows our decision landscape in undesirable and unpredictable ways, including further decisions about whether to use such systems. This dynamic could undermine both user agency and autonomy.

Particularly if decision aid systems are given a centrally determined set of criteria that are applied for conditioning what is presented to the user (e.g. the kind of conditioning that makes nudges possible), we also create an incentive for people to try to exploit those criteria for their own purposes. Similar attack surfaces can appear if models are simply very highly correlated due to being trained on the same datasets. We can already see this at work in agencies commercially offering search engine optimization services, which have expanded to include optimization for appearing in ChatGPT responses. The more criteria are centrally set, the greater the attack surface for those efforts, and the more drastic their potential consequences.

Doing well in life in the coming decades will likely require making more deliberate decisions than it did in previous decades. AI decision aids may be inevitably needed to navigate this world. On the other hand, as just noted, particularly if we view the role of AI decision aids through the framework of nudges, it looks like AI decision tools come with serious risks of making us worse off, even if they are in principle designed with good intentions. I suggest that the central problem that needs to be resolved is how to create AI decision aids that do not inhibit decentralized adaptive learning, and that augment rather than inhibit the agency and autonomy of the user.

\section{Philosophy as the paradigm of autonomy-preserving change in view}

We need to consider what design criteria are necessary to ensure that personal AI systems can enhance both agency and autonomy, in a way that supports rather than inhibits decentralized adaptive learning. These criteria aim to address the challenges and risks outlined in the previous sections while capitalizing on the potential of AI as a tool for supporting human decision-making. Below, I propose a set of design features for a new generation of personal AI systems.

The central problem is that these systems have to be powerful enough to be potentially involved in changing our views about almost anything, or at least very wide range of issues, while supporting rather than undermining our autonomy. There are some early signs that it is not too ambitious a goal to achieve.\footnote{Thomas H. Costello et al., Durably reducing conspiracy beliefs through dialogues with AI. \textit{Science} 385, eadq1814 (2024). DOI: 10.1126/science.adq1814} What I mean by autonomy is a form of ``self-rule'' or ownership of one's judgments, both about what to do and about what it is the case. I take as given that there is such a thing as some judgments being ours, and I will be interested in what kinds of factors are autonomy preserving in this sense. I will call a particular judgment someone might have at a particular point in time, about what is the case or about what to do, a view. Many external impacts on us can engender a change in view in a way that is autonomy preserving. To take a trivial example, you may wonder if it is raining, and change your view to ``it is raining!'' as a result of ordinary perceptual experience. Our interaction with AI systems or with other people also leads to changes in view. The scope of views potentially affected is wider than in the case of ordinary perception. If we use AI to aid us directly in decision making, it not only affects our views about what is the case but also our views about what to do. We then have to ask, what is compatible with a changed view remaining your view rather than an externally imposed view or a view that you are in some sense alienated from? Social science has focused on frameworks where individual preferences are treated as fixed or possibly even identified directly with what is revealed in choice. These views have trouble recognizing autonomy for agents that do not satisfy rationality axioms, since they basically aren't agents at all according to the definition.\footnote{Sen, Amartya. ``Rational Fools: A Critique of the Behavioral Foundations of Economic Theory.'' \textit{Philosophy \& Public Affairs} 6, no. 4 (1977): 317--344.}

The main paradigms for systematic change in view impacted by the influence of intelligence ``from the outside'' have been familiar since the ancients. Rhetoric aims to change views by convincing an audience for a given purpose. We can think of rhetoric as the art of changing views in service of practical aims, in a way that is indifferent to whether this helps in the pursuit of truth. With modern-day recommender systems that underlie social media, shopping, and entertainment platforms, we have found ways to fully automate a very significant dimension of rhetoric in this sense. Plato's commentary on rhetoric in the Georgias, noting that rhetoric is like a kind of ``cookery'' that offers pleasant sensations without guaranteeing nourishment\footnote{Plato. \textit{Gorgias}. Translated by Terence Irwin. Oxford: Clarendon Press, 1979. (465c)}, presages stock criticisms that are routinely levelled against such platforms. Digital rhetoric in this sense is centered on harnessing momentary desires to maintain engagement for its own sake, with changes in view a by-product of this harnessing. However, unless we take a very impoverished notion of a person, the locus of ownership of our views is not well understood as just the stream of momentary desires. Thus, digital rhetoric as a technology for engendering changes in view tends to undermine ownership of our views and our autonomy.

To date there appears to be only one general paradigm of autonomy-preserving change in view: It can be found at the core of philosophy as a practice that traces back to Socrates and is alive and well in many contemporary analytical philosophy departments. Philosophy, in contrast to rhetoric, aims first at seeking wisdom. In areas in which knowledge is possible, it aims at seeking truth. As a practice, it is a form of and contributor to human flourishing. In that sense, it arguably offers ``nourishment'' unlike pure rhetoric. The way philosophy gets to be a contributor to human flourishing is by developing a sustained practice out of something that is already a crucial part of our nature. We already naturally seek truth to a significant extent, we normally prefer reality over fiction, and this is generally conducive to our survival.

Philosophical practice contains within it the insight that change in view is possible in a way that is owned by those whose views are changing and that contributes to truth seeking, and that such a change in view can nonetheless by catalyzed by intelligent inputs from without. This is made possible by the fact that taking on board questions can change your judgment by your own lights, regardless of where the question came from. To illustrate this, let us briefly consider an example that is familiar from virtually any introductory philosophy course. Suppose I hold the view that knowledge should be analysed as justified true belief and that looking at my watch is a good source of justification for beliefs about what time it is. Now suppose I seriously take on board the question of whether we have knowledge in a case in which it is noon, we believe that it is noon as the result of us seeing that our watch says it's noon, but unbeknownst to us, the watch is broken. Most people conclude we do not have knowledge in this case, and as a result they are likely change their view about the idea that knowledge is justified true belief. Taking on board a question about so-called Gettier cases like the one just described has a strong potential to change people's views about knowledge.\footnote{Edmund L. Gettier (1963), ``Is Justified True Belief Knowledge?'', \textit{Analysis}, 23(6), pp. 121--123.} Classic Socratic dialogues are full of similar examples about other concepts. The details of any particular example are secondary. The key feature for our purposes is that here I have ownership of my change in view, regardless of whether the question that prompted this change of view arose in my mind spontaneously, or whether it was given to me by Bertrand Russell\footnote{Bertrand Russell (1948), \textit{Human Knowledge: Its Scope and Limits}}, Edmund Gettier, or indeed Socrates.

We need to dig a bit deeper into what is going well in the foregoing example to draw general lessons. We can of course put emotionally manipulative rhetoric in the form of questions, so it is not the grammatical form of interrogative utterances that has a special status for our purposes. Such form is neither necessary nor sufficient for our purposes. Issues that change our view in the relevant sense could also be raised by looking at something that does not fit our preconceptions, for example. An issue raised indirectly through an observation does not have to be any less conserving of our ownership of our views as the philosophical toy example just discussed. In \textit{Reason and Inquiry}, I argued that a core built-in aim of human reason is to weakly seek erotetic equilibrium.\footnote{Koralus, Philipp. \textit{Reason and Inquiry: The Erotetic Theory}. Oxford: Oxford University Press, 2023.} Roughly, this view holds that we aim to be in a state in which our judgments remain available to us by our own lights regardless of which of an appropriate range of questions we might take on board. In other words, we aim to be robust in our judgment with respect to an appropriate range of questions (or robust with respect to an adequate imaginary range of Socratic interlocutors, if you prefer). On this view, insofar as raising a certain issue furthers our goal of finding this kind of equilibrium, we own the result of having taken on board this question, regardless of whether the question is occurring to us spontaneously or whether it is supplied by someone else, or by an AI system. The key is whether what is supplied to us in fact helps our ability to seek this kind of erotetic equilibrium.

Two important details need to be kept in mind in making sense of the view being suggested here. The first concerns what it means to seek erotetic equilibrium weakly and how it relates to the preservation of autonomy. The second concerns what an ``appropriate'' range of questions is. I will begin with the former. Whether we are in erotetic equilibrium or not does not constitutively depend on how many questions, if any, we have actually tried to take on board. It could be the case that our view would stay the same regardless of what questions we take on board, even though we have not in fact taken any on board. In the general case, we can't just make a policy of taking on board questions willy nilly, since that quickly becomes intractable. An unbounded attempt to take on board questions leads to scepticism and decision paralysis. What I mean here by weakly seeking erotetic equilibrium is that at a minimum, we are responsive to questions when they do arise, should they change our views. In practice, there are significant limits to the range of questions we can handle without cluttering our minds. As a result, we need to make an ``erotetic home'' for ourselves, where we can generally expect that taking direct answers to our questions will tend to yield equilibrium views, even without extensive further inquiry. If I have stocked my fridge with care, I can more or less directly answer my question ``what should I eat?'' by looking inside the fridge, and expect that what I end up choosing would remain the same even if I were to take on board a wide range of questions about the health-related properties of my options, the trade-offs involved in going shopping instead, and so on. In other words, my default judgment is very likely to be in erotetic equilibrium in this place I have made my home. By contrast, if my fridge is primarily stocked with cake and champagne, my default judgment is unlikely to be one that would survive the sorts of questions that I take myself to be on the hook for when it comes to dietary choices. In this sense, a bachelor pad fridge is not an erotetic home for me, since the decisions that come to me most easily in that environment are not likely to be decisions that I would stick to if I took time to reflect.

Since our decisions are sensitive to the questions we have on board, and because our ability to reach equilibrium judgments via tractable ways of answering our questions depends on context, it is not generally the case that decision questions that are equivalent in the objective options they envisage will yield the same judgment of what to do. Changing a decision question in a way that conserves objective options does not guarantee that erotetic equilibrium is conserved. As a result, a nudge intervention is at risk of undermining our autonomy, as it potentially interferes with our strategies for seeking erotetic equilibrium in a way that remains tractable for us. For example, when faced in a hurry with the practically unbounded menu options of a food delivery service in a major metropolis, I might default to ordering fish and vegetables from the nearest high-quality restaurant on the map. If I were to stop and reflect deeply on my choices first, I would have a strong tendency to end up with the same choice, since I would consider the benefits of high-protein diets with vegetables, the importance of a short delivery journey, etc. Now, a platform designer might take the view that alerting customers to especially popular items (``Have you seen what's most popular today in your area?'') is a legitimate way of making their customers more satisfied. After all, restaurant popularity harnesses the experience of many individuals and helps address the fact that the platform cannot itself guarantee that all listed restaurants are good. However, if my display is emblazoned with a big sign alerting me to what is most popular, I might start out with a pang of worry that I am missing out on something that everyone else is in on, and let that impulse determine my choice. After all, I'm in a hurry. However, ``fear of missing out'' is not a criterion of choice that would survive further inquiry, and I would likely not end up with a judgment that is in erotetic equilibrium, since what is most popular is unlikely to correspond to the criteria that are most important to me if I take the time to reflect.

With the above case in mind, we can observe that an external party merely asking questions or merely highlighting information that is already available to me does not in itself guarantee that an agent's resulting change in view is autonomy preserving. The sophist can superficially resemble the philosopher. We additionally need that the question-raising is aimed at supporting rather than undermining the agent's pursuit of erotetic equilibrium. This means that the support system, or our philosophical interlocutor herself, needs to be aiming at erotetic equilibrium as well, relative to an implicitly shared space of questions. In practice, in order for a system to support us in truth seeking, it has to be in some appropriate sense truth seeking as well. This means that, since it has to base its support of our truth seeking in its own truth seeking, as an activity that needs enough independence to actually be able to have a ground from which to support us, the system we need really does have to be an agent, or, more precisely, an epistemic agent.

The second key detail to consider is what an appropriate range of questions consists in. Ultimately, this (meta) question itself can never be answered fully, since it would have to be settled by a meta-inquiry. Yet, the boundaries of such a meta-inquiry are queried within that inquiry itself. Generally speaking, our judgments should be in equilibrium with respect to a wide range of questions, to avoid dogmatism and inability to adapt. However, there are limits that can be delineated with respect to what questions we are not generally on the hook for. If we are asking questions in the domain of empirical science, the practice of empirical science rules many questions out of bounds for its purposes. For example, the question of whether our observations might be the result of an evil demon trying to deceive us, and how we should deal with that in the analysis of our experiments is not one with respect to which we generally demand erotetic equilibrium. Put differently, normal scientific practice does not depend on first resolving philosophical scepticism. Scientists do not proceed with their practice because they have an answer to the sceptic but because they consider the sceptic irrelevant to their practice. In the realm of decision-making, we similarly have bounds. If we are currently having a lovely dinner with friends, it is not a requirement of reason to seek equilibrium with respect to the question of whether there is another dinner table with even lovelier friends we could be attending instead. This remains true even if we would be able to recognize the second table as better, were we to take on board the question. This perspective allows us to reconcile the intuition that it is not irrational to be satisficing instead of optimizing\footnote{Simon, Herbert A. ``Rational Choice and the Structure of the Environment.'' \textit{Psychological Review} 63, no. 2 (1956): 129--138.} with the intuition that it would be irrational to refuse to recognize an option as better when we are directly confronted with it. Seeking erotetic equilibrium weakly as opposed to aggressively does not amount to dogmatism.

\section{The inquiry complex}

If we take the foregoing view seriously, the upshot is this: the paradigm of philosophy offers a path to changes in view that can be impacted by an outside system while preserving autonomy as ownership of our views. This requires that the system helps rather than inhibits our ability to make judgments while weakly seeking erotetic equilibrium with respect to an appropriate range of questions. Appropriate ranges of questions in this sense amount to inquiry complexes that reflect the decentralized adaptive wisdom of communities of truth seekers. Ideally, such a system would turn more of the world into an erotetic home for us, allowing us to reach erotetic equilibrium more efficiently on a wider range of issues that matter to us, without overburdening our cognitive abilities.

Communities of truth-seekers tend to share an evolving joint understanding of what types of questions our judgments ought to be in equilibrium with respect to. This in practice is in itself a form of wisdom subject to decentralized adaptive learning. I call a structured set of questions representing this wisdom at a given stage of adaptation an ``inquiry complex''. Joint truth-seeking of the kind paradigmatic in philosophy can be seen as proceeding against the background of an evolving inquiry complex in this sense. The practice of philosophy shows that it is quite natural for an implicit consensus to emerge about what questions are important to consider in an area, even among practitioners who do not agree on key substantive judgments. As a result, philosophers who have very different views often nonetheless make very similar judgments about whether a given piece of work is of acceptable quality, since the central implicit standard is based on whether the work seems shows an appropriate consideration of the range of central issues around the topic being discussed. An autonomy preserving AI assistant would have to embody inquiry complexes for a broad range of subject matters, mediated through an ecosystem of digital communities.

The goal should be to design a system that in many respects mirrors the performance of a good philosophy tutor (though without limiting topics to those of philosophy as an academic subject). A good philosophy tutor is not in the business of trying to convince people of particular philosophical views. Rather, the philosophy tutor helps the student with two things: make sure the student has awareness of central views and questions that pertain to the general subject of the set of tutorials and help the student navigate those questions to arrive at a considered view. Such a view at a minimum has to be in equilibrium with respect to the questions that are central to the problem space being discussed. The tutor's job is to make sure the student does not just parrot a view or hold some naïve view dogmatically. The tutor's own view of what the right view in the area is will certainly guide what particular range of questions will be explored, but this will not reflect a direct attempt to convince the student of the tutor's view. In fact, to a relative novice who has not read the tutor's own writing, it may often be entirely unclear what view the tutor in fact has. The functional role of the tutor's own view in the area is to provide the basis of a model of a route to erotetic equilibrium in the area. At a minimum the tutor will make sure the student takes the key questions on board that overall in the tutor's view are best answered by the view the tutor holds. But this is entirely compatible with the student answering those questions differently. Yet, the effect is that the student will not end up holding a view that by the student's own lights is only tenable if we do not consider those questions.

Individuals will differ to some extent in their answers to key questions that hinge on intuitions that are hard to break down further (e.g. is this an example of a morally right action? Is this an example case of knowledge?, etc.). As a result, this process does not guarantee that the student ends up with the same conclusion as the tutor. Moreover, it is not the case that nothing has been gained or learned if the tutor and the student don't arrive at the same view. Both may well end up being in erotetic equilibrium. Philosophy may well be the only strongly analytic discipline in which substantial disagreement on the subject matter of a dissertation between a student and an advisor is neither a sign that the advisor/advisee relationship does not work nor in itself a reason to think that the student will not have a career. What I suggest is that we need AI agents that can help us with questions on a much broader basis not limited to what is narrowly considered philosophy, but using the same dynamics of inquiry. The first quarter of the 21st century gave us automated rhetoric. Let the aim for the coming years be to find the right framework for automating philosophy.

\section{Design aspects of AI systems to support decentralized truth seeking}

In this section, we will change gears and consider in outline what aspects an AI system would need to have in order to enable the paradigm we considered in the previous section. The most central aspects for this paradigm have to do with inquiry complexes and seeking erotetic equilibrium.

\textbf{Seeking erotetic equilibrium.}  The system should be aiming to help the user efficiently reach erotetic equilibrium relative to the given set of inquiry complexes, while minimizing the cognitive burden.

\textbf{Individualization of inquiry complexes.} The system must be highly personalized in a way that makes outputs grounded in the user's inquiry complexes for the range of issues that the system will get involved in. This includes being able to integrate evolving inquiry complexes from those communities of truth-seekers that the individual user wishes to join.

If we want the system to be supporting us across potentially all subjects, which would be necessary to address the dilemma from the introduction, we have to recognize that the system will be capable of impacting our choices on core aspects of our lives that are part of what defines who we are. In the previous section, I argued that our paradigm for how such impact can happen is autonomy preserving. Yet, we would then still have to make sure that the implementation of the system does not open up indirect threats to autonomy. This makes privacy, security, and control extremely important.

\textbf{Privacy as a Cornerstone.} A system that helps augment a core cognitive function like truth seeking inquiry across all domains needs to operate at a level of privacy fit for protecting freedom of thought, an even higher standard than is needed for freedom of expression. Privacy must be embedded in every aspect of the system's design to ensure users can engage with the AI without fear of invasion of privacy and the threat of self-censorship at the level of individual deliberation.

\textbf{Security and Trustworthiness.} Trust is essential for users to rely on AI systems that help them seek erotetic equilibrium in significant decision-making contexts. These systems must be resilient against injection and spoofing, and ensure that users do not need to second-guess their outputs.

\textbf{Decentralized Control and Ownership.} Users must be able to retain ownership over their AI systems, particularly regarding critical aspects of configuration and functionality of inquiry complexes. This includes preventing unauthorized changes, ensuring continuity of service, and avoiding vulnerabilities to corporate or governmental coercion. Systems must be designed to empower users, not make them dependent on external actors for critical decisions. By supporting decentralized development and deployment, vulnerabilities to manipulation at the societal level can be reduced.

Finally, in order for our system to be able to remain most effective in the long run, it needs to be able to adapt to changing circumstances and new innovations.

\textbf{Mutual educability.} AI systems should not operate in isolation. They must be designed to learn from users and from other AI systems in ways that users authorize, enabling collaborative and adaptive evolution, through pooling of inquiry complexes.

\textbf{A marketplace of agents and inquiry complexes.} There should be a robust ecosystem where AI systems can compete on meaningful dimensions, such as effectiveness, alignment with user personality, and ease of use. A marketplace for these systems as well as for the inquiry complexes these systems can recruit would encourage innovation and make the overall community of AI aided truth-seekers more adaptive.

\textbf{Modularity for Innovation.} While private ownership of a core is vital, this must not preclude the system's ability to benefit from ongoing technical advancements in other products. A modular design would allow users to integrate new features, improvements, ensuring that their system remains up-to-date and effective without compromising autonomy or security. This may mean separating inquiry complex storage from inference processing, and new concepts for privacy preserving inference requests to external systems, where privacy does not depend on a trusted third party.

Designing a system that can give due regard to the above aspects poses significant conceptual and technical challenges. Yet, there seems to be no reason in principle to suppose that these cannot be solved.

\section{Conclusion}

A core dilemma shapes our digital age: as AI systems grow more powerful and life's choices more complex, we risk either surrendering our agency (becoming overwhelmed by decisions we can no longer make effectively) or relinquishing our autonomy (letting external systems quietly engineer our choices according to priorities not our own). This tension between losing one's grip on the world, through ineffectiveness or cognitive overload, and losing one's self-rule through subtle manipulation, defines the treacherous passage modern technology has opened for us.

The nudge framework was once heralded as a way to steer behavior toward the good without force. But at AI-driven scale, nudging easily stretches from soft paternalism into soft totalitarianism. If we can ``architect'' every context of choice in a fine-tuned and personalized way, a small group of designers or agencies can invisibly shape our decisions. What was meant to sustain our agency risks co-opting it and stifling adaptive learning.

The central question then becomes how to develop a new form of AI; one that avoids the traps of the nudge paradigm and strengthens both our ability to act and our capacity for self-rule. This paper argued for a philosophic turn for AI. Rather than automating rhetorical techniques that funnel users toward outcome templates, we can design truth-seeking agents: systems that expose us to key questions and more efficiently help us reach erotetic equilibrium. By aligning technology with the practice at the core of philosophical inquiry and decentralized adaptation, we can provide a path for individuals to avoid a rigid world of optimized defaults. Instead, we foster an environment where each individual's judgment remains genuinely their own, yet is equipped with the computational power to handle unprecedented complexity.

\end{document}